# Hidden Multiplicity in Exploratory Multiway ANOVA: Prevalence and Remedies

Angélique O. J. Cramer[1*], Don van Ravenzwaaij[2], Dora Matzke[1], Helen Steingroever[1], Ruud Wetzels[3], Raoul P. P. P. Grasman[1], Lourens J. Waldorp[1], Eric-Jan Wagenmakers[1]

[1] Psychological Methods, Department of Psychology, University of Amsterdam, the Netherlands

[2] Faculty of Science and Information Technology, School of Psychology, University of Newcastle, Australia

[3] Data Analytics, PriceWaterhouseCoopers

[*] Corresponding author

E-mail: aoj.cramer@gmail.com

# Abstract

Many psychologists do not realize that exploratory use of the popular multiway analysis of variance (ANOVA) harbors a multiple comparison problem. In the case of two factors, three separate null hypotheses are subject to test (i.e., two main effects and one interaction). Consequently, the probability of at least one Type I error (if all null hypotheses are true) is 14% rather than 5% if the three tests are independent. We explain the multiple comparison problem and demonstrate that researchers almost never correct for it. To mitigate the problem, we describe four remedies: the omnibus $F$ test, the control of familywise error rate, the control of false discovery rate, and the preregistration of hypotheses.

The factorial or multiway analysis of variance (ANOVA) is one of the most popular statistical procedures in psychology. Whenever an experiment features two or more factors, researchers usually apply a multiway ANOVA to gauge the evidence for the presence of each of the separate factors as well as their interactions. For instance, consider a response time experiment with a 2x3 balanced design (i.e., a design with equal number of participants in the conditions of both factors); factor A is speed-stress (high or low) and factor B is the age of the participants (14-20 years, 50-60 years, and 75-85 years). The standard multiway ANOVA tests whether factor A is significant (at the .05 level), whether factor B is significant (at the .05 level) and whether the interaction term A*B is significant (at the .05 level). In the same vein, the standard multiway ANOVA is also frequently used in non-experimental settings (e.g., to assess the potential influence of gender and age on major depression).

Despite its popularity, few researchers realize that the multiway ANOVA brings with it a problem of multiple comparisons, in particular when detailed hypotheses have not been specified *a priori* (to be discussed in more detail later). For the 2x3 scenario discussed above without *a priori* hypotheses (i.e., when the researcher's attitude can be best described by "let us see what we can find"; de Groot, 1969), the probability of finding at least one significant result given that the data originate from the null hypotheses lies in the vicinity of $1 - (1 - .05)^3 = .14$.[1] This is called a Type I error or familywise error rate (FWE). The problem of Type I error is not trivial: add a third, balanced factor to the 2x3 scenario (e.g., a 2x3x3 design), and the probability of finding at least one significant result when $H_0$ is true increases to around 30% ($1 - (1 - .05)^7$), the precise probability depending on to what extent the tests are correlated (see also Footnote 1). Thus, in the absence of strong *a priori* expectations about the tests that are relevant, this alpha-inflation can be substantial and cause for concern.

---

[1] The probability of finding at least one significant result equals exactly 14% iff the three tests are completely independent. This is only true if the total number of participants in the sample approaches infinity: in that case, the *F*-tests become asymptotically independent. For all other sample sizes, the test statistics are not independent because they share a common value, namely the mean square error in the denominator (Feingold & Korsog, 1986; Westfall, Tobias & Wolfinger, 2011). This induces dependence among the test statistics. Another way in which dependence between the tests is induced is when the design is *unbalanced*, i.e., with unequal numbers of participants per condition. The consequence of the dependence between the test statistics is that the probability of finding at least one significant result, given that all null hypotheses are true, will be slightly lower than 14%.

Here we underscore the problem of multiple comparisons inherent in the exploratory multiway ANOVA. We conduct a literature review and demonstrate that the problem is widely ignored: recent articles published in six leading psychology journals contain virtually no procedures to correct for the multiple comparison problem. Next we outline four possible remedies: the omnibus F test, the control of familywise error rate using the sequential Bonferroni procedure, the control of false discovery rate using the Benjamini-Hochberg procedure, and the preregistration of hypotheses.

# Background: Type I Errors and the Oneway ANOVA

A Type I error occurs when a null hypothesis ($H_0$) is falsely rejected in favor of an alternative hypothesis ($H_1$). With a single test, such as the oneway ANOVA, the probability of a Type I error can be controlled by setting the significance level $α$. For example, when $α = .05$ the probability of a Type I error is 5%. Since the oneway ANOVA comprises only one test, there is no multiple comparison problem. It is well-known, however, that this problem arises in the oneway ANOVA whenever the independent variable has more than two levels and post-hoc tests are employed to determine which condition means differ significantly from one another. For example, consider a researcher who uses a oneway ANOVA and obtains a significant effect for Ethnicity on the total score of a depression questionnaire. Assume that Ethnicity has three levels (e.g., Caucasian, African-American, and Asian); then this researcher will usually perform multiple post-hoc tests to determine which ethnic groups differ significantly from one another – here the three post-hoc tests are Caucasian vs. African-American, Caucasian vs. Asian, and African-American vs. Asian. Fortunately, for the oneway ANOVA the multiple comparison problem has been thoroughly studied. Software programs such as SPSS and SAS explicitly address the multiple comparison problems by offering a host of correction methods including Tukey's HSD test, Hochberg's GT2, and the Scheffé method (Hochberg, 1974; Scheffé, 1953; Tukey, unpublished; Westfall, Tobias & Wolfinger, 2011).

# The Exploratory Multiway ANOVA: A Family of Hypotheses

Now consider a design that is only slightly more complicated. Suppose a researcher wants to test whether both Gender (G; two levels) and Ethnicity (E; three levels) influence the total score on a depression questionnaire. Furthermore, suppose that this researcher has no firm *a priori* hypothesis about how G and E influence the depression total score; that is, the researcher is predominantly interested in finding out whether *any* kind of relationship exists between G, E and depression: a classic example of the *guess* phase of the empirical cycle in which hypotheses are formed rather than tested (de Groot, 1969).

In this case, the multiway ANOVA with two factors, G and E, is an *exploratory* one: without strictly formulated *a priori* hypotheses, the researcher obtains the results for all three hypotheses involved (i.e., main effect of G, main effect of E and a GxE interaction) by means of a single mouse click in SPSS. As such, in an exploratory setting, all hypotheses implied by the design are considered and tested jointly, rendering this collection of hypotheses a *family*; in line with the idea that "…the term 'family' refers to the collection of hypotheses […] that is being considered for joint testing" (Lehmann & Romano, 2005). As a result, we argue that a multiple comparison problem lurks in these exploratory uses of a multiway ANOVA.

To see this, consider the results of a fictitious exploratory multiway ANOVA as reported in Table 1. When interpreting the ANOVA table, most researchers would conclude that both main effects as well as the interaction are significant as all *p*-values are smaller than $\alpha = .05$. This conclusion is intuitive and directly in line with the numbers reported in Table 1. Nevertheless, this conclusion is statistically unwarranted; the researcher does not have firm *a priori* hypotheses and therefore tests all three hypotheses simultaneously, engaging in an exploratory research effort. In this case, when all null hypotheses are true, the Type I error will be larger than 5% (around 14%, see Footnote 1). Note that multiway ANOVAs in the psychological literature often consist of three or four factors and this compounds the problem. In the case of three factors, without *a priori* hypotheses, and when all null hypotheses are true, the total number of tests is seven (i.e., three main effects, three first-order interactions, and one second-order interaction, $2^3 - 1$) and the resulting probability of a Type I error around 30% (i.e., $1 - (1 - .05)^7$); with four factors and when all null hypotheses are true, the probability of incorrectly rejecting one or more null hypotheses is around 54%. It is therefore incorrect to compare each of the *p*-values from a multiway ANOVA table to $\alpha = .05$.

This is notably different from the situation where the researcher uses a multiway ANOVA for *confirmatory* purposes; that is, where the researcher tests one or more *a priori* postulated hypotheses (i.e., hypothesis testing in the *predict* phase of the empirical cycle; de Groot, 1969). In the case of one predefined hypothesis in a design with two factors, for example, the family is no longer defined as encompassing all hypotheses implied by the design (i.e., three); but as all to-be-tested hypotheses, in this case: one, rendering it unnecessary to adjust the level of $α$.

The realization that exploratory multiway ANOVAs inherently contain a multiple comparison problem may come as a surprise to many empiricists, even to those who use the multiway ANOVA on a regular basis. In standard statistical textbooks, the multiple comparison problem is almost exclusively discussed in the context of one-way ANOVAs (with Westfall, Tobias & Wolfinger, 2011, as notable exception). In addition, statistical software packages such as SPSS do not present the possible corrective procedures for the multiway case, and this invites researchers to compare each of the p-values to $α = .05$.

We are not the first to identify the multiplicity problem in the multiway ANOVA (e.g., Didelez, Pigeot & Walter, 2006; Fletcher, Daw & Young, 1989; Kromrey & Dickinson, 1995; Olejnik & Supattathum, 1997; Ryan, 1959; Smith, Levine, Lachlan & Fediuk, 2002). Earlier work on the problem, however, does not feature in mainstream statistical textbooks. Moreover, the majority of this work is written in a technical style that is inaccessible to scholars without sophisticated statistical knowledge. Consequently, empirical work has largely ignored the multiplicity problem in the multiway ANOVA. As we will demonstrate shortly, the ramifications can be profound.

One may argue that the problem sketched above is less serious than it appears. Perhaps the majority of researchers in psychology test a single pre-specified hypothesis, thereby circumventing the multiple comparison problem. Or perhaps, whenever they conduct multiple tests, they use some sort of procedure to adjust the $\alpha$ level for each test. This is not the case. Pertaining to the former, it is unfortunately quite common to perform what Gigerenzer (2004) has termed the "null ritual" in which a researcher specifies $H_0$ in purely statistical terms (e.g., equality of means) without providing an alternative hypothesis in substantive terms (e.g., women are more depressed than men). Additionally, Kerr (1998) notes that researchers in psychology are quite commonly seduced into presenting a post hoc hypothesis (e.g., Caucasian people are more depressed than African-American people: main effect of ethnicity on depression) as if it were an *a priori* hypothesis (Hypothesizing After the Results are Known: HARKing; see also Barber, 1976). Hence, hindsight bias and confirmation bias make it difficult for researchers to ignore the presence of unexpected "significant" effects (i.e., effects for which the individual test has $p < .05$).

The next section addresses the empirical question of whether researchers correct for multiple comparisons when they use the multiway ANOVA. The short answer is that, almost without exception, researchers interpret the results of the individual tests in isolation, without any correction for multiple comparisons.

# Prevalence: Multiway Corrections in Six Psychology Journals

We selected six journals that rank among the most widely read and cited journals in experimental, social, and clinical psychology. For these journals we specifically investigated all 2010 publications:

1. *Journal of Experimental Psychology General*: volume 139, issues 1-4 (40 papers).

2. *Psychological Science*: volume 21, issues 1-12 (285 papers).

3. *Journal of Abnormal Psychology*: volume 119, issues 1-4 (88 papers).

4. *Journal of Consulting and Clinical Psychology*: volume 78, issues 1-6 (92 papers).

5. *Journal of Experimental Social Psychology*: volume 46, issues 1-6 (178 papers).

6. *Journal of Personality and Social Psychology*: volumes 98 and 99, issues 1-6 (136 papers).

For each article, we assessed whether a multiway ANOVA was used. If so, we investigated whether the authors had used some sort of correction procedure (e.g., an omnibus test) to remedy the multiple comparison problem. The results are summarized in Table 2.

Two results stand out. First, almost half of all articles under investigation here used a multiway ANOVA, underscoring the popularity of this testing procedure. Second, only around 1% of these papers used a correction procedure (i.e., the omnibus *F*-test, see below).

In sum, our literature review confirms that the multiway ANOVA is a highly popular statistical method in psychological research, but that its use is almost never accompanied by a correction for multiple comparisons. Note that this state of affair is different for fMRI and genetics research where the problem is more evident and it is common practice to correct for multiplicity (e.g., Poldrack et al., 2008).

# Remedies

As noted earlier, some statisticians have been aware of the multiple comparison problem in multiway ANOVA. However, our literature review demonstrated that this awareness has not resonated in the arena of empirical research in psychology. Below we discuss four different procedures to mitigate the multiple comparison problem in multiway ANOVA: (1) the omnibus *F* test; (2) controlling the familywise error rate; (3) controlling the false discovery rate; and (4) preregistration.

## Remedy 1: The Omnibus *F* Test

In the few cases where a correction procedure was used, this involved an omnibus *F* test. In such a test, one pools the sums of squares and degrees of freedom for all main effects and interactions into a single *F* statistic. The individual *F* tests should only be conducted if this omnibus $H_0$ is rejected (Fletcher, Daw & Young, 1989; Wright, 1992). So for example, in the case of a 2x2 ANOVA, one should first test the omnibus hypothesis with all three hypotheses included (two main effects and an interaction). If significant, one may continue and test the individual hypotheses.

However, the omnibus *F* test does not control the familywise Type I error under partial null conditions (Kromrey & Dickinson, 1995). For example, suppose that in a three-way ANOVA a main effect is present for one factor but not in the remaining two factors; then the overall *F* test is likely to yield a significant *F* value because, indeed, the omnibus null hypothesis is false. However, the omnibus test does not remedy the multiple comparison problem involving the remaining two factors. Hence, the omnibus *F* test offers only weak protection against the multiplicity problem.

## Remedy 2: Controlling Familywise Error Rate

The familywise error rate (FWER) refers to the probability of making at least one Type I error within the family of tests under consideration; here the family consists of all tested effects in a multiway ANOVA without *a priori* hypotheses. To control this FWER one has to make certain that it is smaller than or equal to $\alpha$, which usually equals 5%. Preferably, FWER is controlled in the *strong sense,* such that it holds for any configuration of true and false null hypotheses.

One method to control FWER in the strong sense is the sequential Bonferroni procedure (also known as the Bonferroni-Holm correction), which was first introduced by Hartley (1955) and subsequently (independently) re-invented and/or modified by others (Hochberg, 1988; Holm, 1979; McHugh, 1958; Rom, 1990; Shaffer, 1986; Wright, 1992). To illustrate the procedure, let us revisit our hypothetical example in which a researcher conducts a two-way ANOVA with G and E as independent factors (uncorrected results are listed in **Table 1**). The results of the sequential Bonferroni correction procedure for this example are presented in **Table 3**. First, one sorts all significant *p*-values in ascending order, that is, with the smallest *p*-value first (see also **Figure 1** for a visual explanation of the method). Next, one computes an adjusted *α* level, $α_{adj}$. For the smallest *p*-value, $α_{adj}$ equals *α* divided by the number of tests. Thus, in this example, we conduct three tests so $α_{adj}$ for the smallest *p*-value equals .05/3 = .01667. For the second *p*-value, $α_{adj}$ equals *α* divided by the number of tests minus 1. So, in our example, the next $α_{adj}$ equals .05/2 = .025. For the final *p*-value, $α_{adj}$ equals *α* divided by 1 (i.e., the total number of tests minus 2). So, in our example, the final $α_{adj}$ equals .05/1 = .05. Next, one evaluates each *p*-value against these adjusted *α* levels, sequentially, with the smallest *p*-value evaluated first. Importantly, if the $H_0$ associated with this *p*-value is not rejected (i.e., $p > α_{adj}$) then all testing ends and all remaining tests are considered non-significant as well.

***************************************

PLACE FIGURE 1 ABOUT HERE

***************************************

In our example, we evaluate $p$ = .0195 against $α_{adj}$ = .01667: $p > α_{adj}$ and therefore we conclude that the G x E interaction is not significant. The sequential Bonferroni procedure mandates that we stop testing and conclude that the remaining main effects are not significant either. Thus, when the sequential Bonferroni correction procedure is applied to our example, none of the effects are significant; without a correction procedure, all of the effects are significant.

Thus, the sequential Bonferroni correction procedure allows control over the FWE by evaluating each null hypothesis – from the one associated with the smallest to the one associated with the largest $p$-value – against an $α$ level that is adjusted in order to control for the inflated probability of a Type I error. In this way, the probability of rejecting one or more null hypotheses while they are true will be no larger than 5% (for a proof see Hartley, 1955). Note that for relatively small number of tests $k$, the sequential Bonferroni correction is notably less conservative than the standard Bonferroni correction where one divides $α$ by $k$ for all null hypotheses. However, sequential Bonferroni is still a relatively conservative procedure in that it always retains the remaining $H_0$ whenever one $H_0$ is not rejected, regardless of how many remain. That is: it does not matter whether one has five or 50 null hypotheses, one single $H_0$ that is not rejected means that all remaining null hypotheses are also not rejected. As such, some have argued that procedures such as (sequential) Bonferroni, while adequately controlling the probability of a Type I error, reduce power to find any effect and thus inflate the probability of a Type II error (not rejecting $H_0$ while the alternative hypothesis $H_1$ is true; e.g., Benjamini & Yekutieli, 2001; Nakagawa, 2004).

Another disadvantage of the sequential Bonferroni procedure is conceptual: the significance of a particular factor depends on the significance of other, unrelated factors. For instance, the main effect for G reported in Table 1 has *p* = .0329. If the effects for the other two factors (i.e., E*G and E) had been more compelling (e.g., *p* = .01 for both), the final and third test for G would have been conducted at the *α* = .05 level, and the result would have been labeled significant. This dependence on the result for unrelated tests may strike one as odd.

The sequential Bonferroni procedure is by no means the only one in its class, and we present it here merely as a prototypical example of a procedure that seeks to control FWER. A well-known alternative procedure is the regular Bonferroni correction in which α, for every *p*-value alike, is divided by the total number of tests. As such, the regular Bonferroni correction does not have the conceptual drawback of the significance of one result being dependent on the other results for unrelated tests. However, compared to sequential Bonferroni, the regular Bonferroni is inferior in terms of power. Other methods to control FWER are for example the Simes procedure (Simes, 1986) and the Hommel correction (Hommel, 1988).

## Remedy 3: Controlling False Discovery Rate

An alternative might be to forego control of FWER and instead control the *false discovery rate* (FDR; Benjamini, Drai, Elmer, Kafkaki, & Golani, 2001; Benjamini & Hochberg, 1995), which is the expected proportion of erroneous rejections of $H_0$ among all rejections of $H_0$. When controlling FDR, the probability of a Type II error is smaller than controlling FWER but this comes at the expense of a higher probability of a Type I error. Controlling FDR is particularly appropriate for applications in genetics and neuroimaging, where the goal is to identify candidate effects from a large set; these candidates can then be tested more rigorously in follow-up confirmatory experiments.

One way to control FWR is with the Benjamini-Hochberg procedure (BH; Benjamini & Hochberg, 1995). To illustrate the procedure, consider again our hypothetical example for which the uncorrected results are listed in **Table 1**. The results of the BH procedure for this example are presented in **Table 3**, and they were obtained as follows: First, one sorts all *p*-values in ascending order, that is, with the smallest *p*-value first (see also **Figure 2** for a visualization of the method). Next, one computes an adjusted *α* level, $α_{adj}$. For the largest *p*-value, $α_{adj}$ equals $α$ times the rank number of the largest *p*-value (3 in our example) divided by the total number of tests (also 3 in this example): 0.05*(3/3) = 0.05. For the middle *p*-value, $α_{adj}$ equals 0.05*(2/3) = 0.0333; for the smallest *p*-value, $α_{adj}$ equals 0.05*(1/3) = 0.01667. Next, one evaluates each *p*-value against these adjusted *α* levels, with the largest *p*-value evaluated first. Importantly, if the $H_0$ associated with this *p*-value is rejected (i.e., *p* < $α_{adj}$) then all testing ends and all remaining tests are considered significant as well.

****************************************

PLACE FIGURE 2 ABOUT HERE

****************************************

In our example, we evaluate *p* = .0329 against $α_{adj}$ = .05: *p* < $α_{adj}$ and therefore we conclude that the main effect of G is significant (and thus $H_0$ is rejected). According to the BH procedure, we stop testing and conclude that this main effect as well as the other main effect and the interaction are significant. Note that this conclusion is drawn despite the fact that the *p*-value for the G x E interaction exceeded the adjusted alpha level. In the alternative situation that we would have retained the null hypothesis of the first *p*-value, the testing would have continued by evaluating the second *p*-value against its adjusted alpha.

The BH procedure is certainly not the only way to control FDR. Other procedures include the Benjamini-Hochberg-Yekutieli procedure (Benjamini & Yekutieli, 2001), which controls FDR under positive dependence assumptions; and the Efron method (Efron, Storey & Tibshirani, 2001a; Efron, Tibshirani, Storey & Tusher, 2001b) that controls not exactly FDR but local FDR, which is the conditional probability that the null hypothesis is true given the data.

## Remedy 4: Preregistration

Another effective remedy is *preregistration* (e.g., Chambers, 2013; Chambers et al., 2013; de Groot, 1969; Goldacre, 2009; Nosek & Lakens, 2014; Wagenmakers, Wetzels, Borsboom, van der Maas & Kievit, 2012; Wolfe, 2013; for preregistration in medical clinical trials see e.g., www.clinicaltrials.gov). By preregistering their studies and their analysis plan, researchers are forced to specify beforehand the exact hypotheses of interest. In doing so, as we have argued earlier, one engages in confirmatory hypothesis testing (i.e., the confirmatory multiway ANOVA), a procedure that can greatly mitigate the multiple comparison problem. For instance, consider experimental data analyzed with a 2x2x3 multiway ANOVA; if the researcher stipulates in advance that the interest lies in the three-way interaction and the main effect of the first factor, this reduces the number of tested hypotheses from seven to two, thereby diminishing the multiplicity concern.

# Conclusion

We have argued that the multiway ANOVA harbors a multiple comparison problem, particularly when this analysis technique is employed relatively blindly, that is, in the absence of strong *a priori* hypotheses. Although this hidden multiple comparison problem has been studied in statistics, empiricists are not generally aware of the issue. This point is underscored by our literature review, which showed that, across a total of 819 articles from six leading journals in psychology, corrections for multiplicity are virtually absent.

The good news is that the problem, once acknowledged, can be remedied in one of several ways. For instance, one could use one of several procedures to control either familywise error rate (e.g., with the sequential Bonferroni procedure) or the false discovery rate (e.g., with the Benjamini-Hochberg procedure). These procedures differ in terms of the balance between safeguarding against Type I and Type II errors. On the one hand, it is crucial to control the probability of rejecting a true null hypothesis (i.e., the Type I error). On the other hand, it is also important to minimize the Type II error, that is, to maximize power (Button et al., 2013). As we have shown in our fictitious data example, towards which side the balance shifts may make a dramatic difference in what one would conclude from the data: when using sequential Bonferroni (i.e., better safeguard against Type I errors at the cost of a reduction in power) all null hypotheses were retained; when using the Benjamini-Hochberg procedure (i.e., less control over Type I errors but more power) all null hypotheses were rejected. So what is a researcher to do when various correction procedures result such different conclusions? It appears prudent to follow the statistical rule of thumb for handling uncertainty: when in doubt, issue a full report that includes the results from all multiple correction methods that were applied. Such a full report allows the reader to assess the robustness of the statistical evidence. Of course, the royal road to obtaining sufficient power is not to choose a lenient correction method; instead, one is best advised to plan for a large sample size (Klugkist, Post, Haarhuis & van Wesel, 2014).

And there is even better news. Many if not all correction methods for controlling either FWER or FDR are easy to implement using the function p.adjust() in the basic *stats* package in R (R Development Core Team, 2007). All that is required is to input a vector of *p*-values, and the function evaluates these according to the chosen correction method.

We realize that our view on differential uses of the multiway ANOVA (i.e., exploratory vs. confirmatory) hinges on the specific definition of what constitutes a family of hypotheses; and we acknowledge that other definitions of such a family exist. However, in our view, the intentions of the researcher (exploratory hypothesis *formation* or confirmatory hypothesis *testing*) play a crucial part in determining the size of the family of hypotheses. It is vital to recognize the multiplicity inherent in the exploratory multiway ANOVA and correct the current unfortunate state of affairs[2]; the alternative is to accept that our findings might be less compelling than advertised.

---

[2] Fortunately, some prominent psychologists such as Dorothy Bishop, are acutely aware of the multiple comparison problem in multiway ANOVA and urge their readers to rethink their analysis strategies: http://deevybee.blogspot.co.uk/2013/06/interpreting-unexpected-significant.html.

Kromrey, J. D., & Dickinson, W. B. (1995). The use of an overall F test to control Type I error rates in factorial analyses of variance: Limitations and better strategies. *The Journal of Applied Behavioral Science*, *31*, 51-64.

Lehmann, E. L., & Romano, J. P. (2005). Generalization of the familywise error rate. *The Annals of Statistics*, *33*, 1138-1154.

McHugh, R. (1958). Significance level in factorial design. *The Journal of Experimental Education, 26*, 257-260.

Nakagawa, S. (2004). A farewell to Bonferroni: The problems of low statistical power and publication bias. *Behavioral Ecology, 15*, 1044-1045.

Nosek, B. A., & Lakens, D. (2014). Registered reports: A method to increase the credibility of published results. *Social Psychology, 45*, 137-141.

Olejnik, S., Li, J., & Supattathum, S. (1997). Multiple testing and statistical power with modified Bonferroni procedures. *Journal of Educational and Behavioral Statistics, 22*, 389-406.

Poldrack, R. A., Fletcher, P. C., Henson, R. N., Worsley, K. J., Brett, M., & Nichols, T. E. (2008). Guidelines for reporting an fMRI study. *NeuroImage, 40*, 409-414.

Rom, D. M. (1990). A sequentially rejective test procedure based on a modified Bonferroni inequality. *Biometrika, 77*, 663-665.
21

# Tables

Table 1. Example ANOVA table for the three tests associated with a hypothetical 2x3 design with Gender (G) and Ethnicity (E) as independent factors.

|  |  | *df1* | *df2* | *F* | *p*-value |
|---|---|---|---|---|---|
| *Main effect* | G | 1 | 30 | 5 | .0329* |
|  | E | 2 | 30 | 4 | .0288* |
| *Interaction* | G x E | 2 | 30 | 4.50 | .0195* |

*significant at *α* = .05

Table 2. Percentage of articles overall and in the six selected journals that used a multiway ANOVA, and the percentage of these articles that used some sort of correction procedure

|  | % papers using mANOVA | % papers using mANOVA + correction |
|---|---|---|
| *Overall* | 47.62 | 1.03 |
| *JEPG* | 84.61 | 0 |
| *Psych Sci* | 43.16 | 0 |
| *J Abn Psych* | 31.82 | 0 |
| *JCCP* | 16.30 | 0 |
| *JESP* | 65.17 | 2.59 |
| *JPSP* | 54.41 | 1.35 |

*Overall*, all papers from the six journals together; *JEPG*, Journal of Experimental Psychology: General; *Psych Sci*, Psychological Science; *J Abn Psych*, Journal of Abnormal Psychology; *JCCP*, Journal of Consulting and Clinical Psychology; *JESP*, Journal of Experimental Social Psychology; *JPSP*, Journal of Personality and Social Psychology; mANOVA, multiway ANOVA.

Table 3. Results from the sequential Bonferroni (*seqB*) and Benjamini-Hochberg (*BH*) procedure for the example from Table 1. $α_{adj}$ *seqB* = the adjusted alpha level with the sequential Bonferroni procedure; $α_{adj}$ *BH* = the adjusted alpha level with the Benjamini-Hochberg procedure; $H_0$ *seqB* = evaluation of the null hypothesis with the sequential Bonferroni procedure; $H_0$ *BH* = evaluation of the null hypothesis with the Benjamini-Hochberg procedure.

| Effect | *p*-value | $α_{adj}$ *seqB* | $α_{adj}$ *BH* | $H_0$ *seqB* | $H_0$ *BH* |
|---|---|---|---|---|---|
| G x E | .0195 | .0167 | .0167 | retained | **rejected** |
| E | .0288 | .0250 | .0333 | retained | **rejected** |
| G | .0329 | .0500 | .0500 | retained | **rejected** |

# Figure Captions

*Figure 1*. A visual representation of the sequential Bonferroni method for controlling familywise error rate. All *p*-values are sorted in ascending order and are assigned a rank number from 1 (smallest) to *k* (largest). Next, one starts by evaluating the first (smallest) *p*-value ($p^{(1)}$) against the adjusted $\alpha$ ($\alpha_{adj}$), which is – for the first *p*-value – equal to $\alpha$ divided by *k*. If the *p*-value is smaller than $\alpha_{adj}$ then the first hypothesis $H^{(1)}$ is rejected and one proceeds to the second *p*-value. If the *p*-value is not smaller than $\alpha_{adj}$ then one immediately accepts all null hypotheses and stops testing.

*Figure 2*. A visual representation of the Benjamini-Hochberg procedure for controlling false discovery rate. All *m* *p*-values are sorted in ascending order and assigned a rank number from 1 (smallest) to *k* (largest). Next, one starts by evaluating the last (largest) *p*-value ($p^{(k)}$) against the adjusted $\alpha$ ($\alpha_{adj}$), which is – for the last *p*-value – equal to *k* divived by *m* times $\alpha$. If the *p*-value is smaller than $\alpha_{adj}$ then all null hypotheses are rejected and testing stops. If the *p*-value is not smaller than $\alpha_{adj}$ then one proceeds to the next *p*-value.

*Figure 1.*

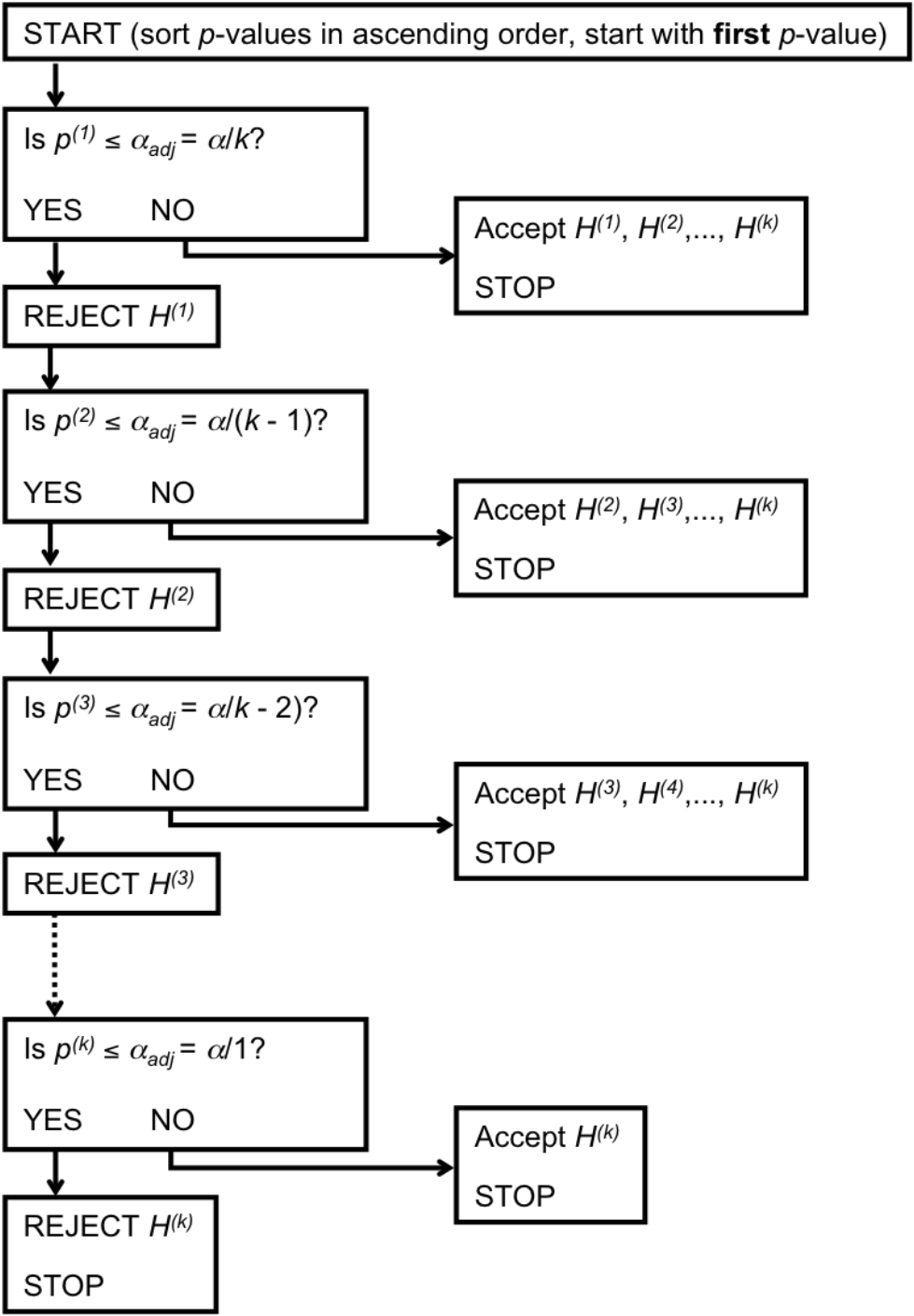